\documentclass[a4paper,twocolumn,10pt]{article}

\usepackage{amssymb,amsbsy,psfig,graphicx,times,bbm}
\newcommand{\gr}[1]{\boldsymbol{#1}}
\newcommand{\be}{\begin{equation}}
\newcommand{\ee}{\end{equation}}
\newcommand{\bea}{\begin{eqnarray}}
\newcommand{\eea}{\end{eqnarray}}

\newcommand{\sirsection}[1]{\section{\large \sf \textbf{#1}}}
\newcommand{\ack}{\section*{\normalsize \sf \textbf{Acknowledgements}}}

\begin{document}

\title{\sf \textbf{\Large Symplectic invariants, entropic measures and correlations\\
of Gaussian states}}
\author{Alessio Serafini, Fabrizio Illuminati 
and Silvio De Siena \vspace*{.4cm}\\
\footnotesize Dipartimento di Fisica ``E.~R.~Caianiello'',
Universit\`a di Salerno, INFM UdR Salerno,\vspace*{-.1cm}\\ 
\footnotesize INFN Sez.~Napoli, 
Gruppo Collegato di Salerno,
Via S. Allende, 84081 Baronissi (SA), Italy}

\date{\footnotesize \sf \begin{quote}
\hspace*{.2cm} We present a derivation of the Von Neumann entropy and mutual information 
of arbitrary two--mode Gaussian states, based on the explicit determination of the symplectic 
eigenvalues of a generic covariance matrix. The key role of the symplectic invariants
in such a determination is pointed out.
We show that the Von Neumann entropy depends on two symplectic invariants, 
while the purity (or the linear entropy) is determined by only one invariant, 
so that the two quantities provide two different hierarchies of
mixed Gaussian states. 
A comparison between mutual information and entanglement of 
formation for symmetric states is considered, 
remarking the crucial role of the symplectic eigenvalues 
in qualifying and quantifying the correlations
present in a generic state.
\end{quote}
December 12, 2003}

\maketitle

\sirsection{Introduction}
Quantum information with continuous variable 
systems is rapidly developing and appears to yield very  
promising perspectives concerning both experimental 
realizations and general theoretical insights.
In such a context, Gaussian states play 
a prominent role, both in view of their conceptual importance
and for their relevance in experimental applications,
and have attracted most of the attention of the researchers 
in the field \cite{pati03}. 
They are the easiest states that can be created and controlled 
in the laboratory \cite{labs}, and have been successfully
exploited in quantum cryptography \cite{crypto} and quantum 
teleportation protocols \cite{tele}. Moreover, they are 
possible candidates for continuous variable quantum computation 
processing \cite{compu,pati03}. \par
In particular, two--mode Gaussian states aroused 
great interest in later years, being the simplest 
prototype of a continuous variable bipartite system.
As for the theory, the qualitative characterization of 
the entanglement of two--mode Gaussian
states has been fully developed by determining
the necessary and sufficient criteria for
their separability \cite{simon,duan}. 
A computable quantitative characterization 
of entanglement for such states is available as well,
being provided by 
negativity \cite{werner} and, 
in the symmetric instance, by the
entanglement of formation \cite{giedke,wolf}.\par
Due to the interaction with the environment,
any pure quantum state involved in some
quantum information process evolves into a mixed state.   
Therefore, another property 
of crucial interest in Quantum Information Theory 
is quantifying the degree of mixedness of a quantum 
state. Let us briefly recall that the 
degree of mixedness of a quantum state 
$\boldsymbol{\varrho}$ can be characterized 
either by the Von Neumann entropy 
$S_{V}({\boldsymbol{\varrho}})$ or by 
the linear entropy $S_{L}({\boldsymbol{\varrho}})$. 
Such quantities are defined as 
follows for continuous variable systems:
\begin{eqnarray}
S_{V}({\boldsymbol{\varrho}})
& \equiv & - \;{\rm Tr}\,({\boldsymbol{\varrho}}\;
\ln{{\boldsymbol{\varrho}}})\; , 
\label{vneu} \\
S_{L}({\boldsymbol{\varrho}}) & \equiv & 1 - 
\;{\rm Tr}\,({\boldsymbol{\varrho}}^{2})
\equiv 1 - \mu({\boldsymbol{\varrho}}) \; ,
\label{linea}
\end{eqnarray}
where $\mu \equiv \, {\rm Tr} \, 
({\boldsymbol{\varrho}}^{2})$ 
denotes the purity of the state ${\boldsymbol{\varrho}}$. 
The linear entropy of an arbitrary $n$--mode 
Gaussian state can be easily computed, whereas
the evaluation of the Von Neumann entropy
requires, in general, a more involved technical
procedure. 
Making use of the Von Neumann entropy is
however preferable, as it allows for a 
finer and more precise characterization of mixedness 
and correlations for multi-mode Gaussian states. 
In fact, the Von Neumann entropy is additive on tensor 
product states, unlike the linear entropy.
Moreover, we will show that for two--mode Gaussian
states the Von Neumann entropy depends on two symplectic 
invariants, while the linear entropy in completely 
determined by only one invariant.
Finally, the knowledge of the Von Neumann
entropy of a generic two--mode state allows to 
obtain the mutual information 
$I({\boldsymbol{\varrho}}) \equiv
S_{V}({\boldsymbol{\varrho}}_{1})
+ S_{V} ({\boldsymbol{\varrho}}_{2})
- S_{V}({\boldsymbol{\varrho}})$ 
(here ${\boldsymbol{\varrho}}_{i}$ 
is the reduced density matrix of 
subsystem $i$), which
quantifies the total amount of correlations 
(quantum plus classical) contained in a 
state \cite{vedral}.\par
The Von Neumann entropy of a $n$--mode Gaussian state 
has been determined in some remarkable works 
on the capacity of Gaussian channels by Holevo {\em et al.} \cite{holevo}.
In this letter we elucidate the mathematical structure 
introduced by these Authors, recasting it in a form
that is conceptually simple and physically useful
for applications and for explicit experimental schemes.
To do so, we provide a thorough and detailed analysis of 
the basics of the symplectic framework
restricting to the two--mode case, presenting
a simple and transparent procedure to
evaluate the Von Neumann entropy and the mutual information 
of two--mode Gaussian states. Our approach is based on the
explicit determination of the global symplectic invariants. 
The methodology we introduce naturally
highlights the role played by the symplectic eigenvalues 
in characterizing the correlations encoded in 
multipartite systems.\par
\sirsection{Two--mode Gaussian states}
Let us consider a two--mode continuous variable system, 
described by the
Hilbert space ${\cal H}={\cal H}_{1}\otimes{\cal H}_{2}$ 
resulting from the tensor product 
of the Fock spaces ${\cal H}_{k}$'s. 
We will call $a_{k}$ the annihilation operator 
acting on the space ${\cal H}_{k}$. 
Likewise, $\hat x_{k}=(a_{k}+a^{\dag}_{k})/\sqrt{2}$ 
and $\hat p_{k}=-i(a_{k}-a^{\dag}_{k})/\sqrt{2}$ are
the quadrature phase operators of the mode $k$,
the corresponding phase space variables being $x_{k}$ and $p_{k}$.
The set of Gaussian states is, by definition, 
the set of states with Gaussian characteristic 
functions and quasi--probability distributions. 
Therefore, a Gaussian state is completely
characterized by its first and second statistical moments, 
that is, respectively, by the vector of 
mean values
$\bar X\equiv\left(\langle\hat x_{1}\rangle,
\langle\hat p_{1}\rangle,\langle\hat x_{2}\rangle,
\langle\hat p_{2}\rangle\right)$ 
and by the covariance matrix $\boldsymbol{\sigma}$
\begin{equation}
\sigma_{ij}\equiv\frac{1}{2}\langle \hat{x}_i \hat{x}_j + 
\hat{x}_j \hat{x}_i \rangle -
\langle \hat{x}_i \rangle \langle \hat{x}_j \rangle \, .
\end{equation}
First moments will be unimportant to our aims, 
and we will set them to zero (as it is always possible 
by means of a local unitary transformation)
without any loss of generality for our results.
For simplicity, in what follows 
$\boldsymbol{\sigma}$ will refer both to the Gaussian state 
and to its covariance matrix.
It is convenient to express 
$\boldsymbol{\sigma}$ in terms of the three $2\times 2$
matrices $\boldsymbol{\alpha}$, $\boldsymbol{\beta}$, 
$\boldsymbol{\gamma}$
\begin{equation}
\boldsymbol{\sigma}\equiv\left(\begin{array}{cc}
\boldsymbol{\alpha}&\boldsymbol{\gamma}\\
\boldsymbol{\gamma}^{T}&\boldsymbol{\beta}
\end{array}\right)\, . 
\label{espre}
\end{equation}
Let us define two further submatrices of 
$\boldsymbol{\sigma}$
\begin{equation}
\boldsymbol{\delta}=\left(\begin{array}{cc}
\sigma_{11}&\sigma_{13}\\
\sigma_{31}&\sigma_{33}
\end{array}\right)\; , \quad 
\boldsymbol{\epsilon}=\left(\begin{array}{cc}
\sigma_{22}&\sigma_{24}\\
\sigma_{42}&\sigma_{44}
\end{array}\right)\; . \label{minors}
\end{equation}
The privileged role played by 
$\boldsymbol{\delta}$ and $\boldsymbol{\epsilon}$ in 
characterizing the action of global symplectic 
operations on $\boldsymbol{\sigma}$
will become clear in the following. \par
Positivity of ${\boldsymbol{\varrho}}$ 
and the commutation relations
for quadrature phase operators impose the following 
constraint ensuring that $\boldsymbol{\sigma}$
be a {\em bona fide} covariance matrix \cite{simon}
\begin{equation}
\boldsymbol{\sigma}+\frac{i}{2}\boldsymbol{\Omega}\ge 0 \; ,
\label{bonfide}
\end{equation}
where $\boldsymbol{\Omega}$ is the usual symplectic form
\begin{equation}
\boldsymbol{\Omega}\equiv \left(\begin{array}{cc}
\boldsymbol{\omega}&0\\
0&\boldsymbol{\omega}
\end{array}\right)\; , \quad \boldsymbol{\omega}\equiv 
\left( \begin{array}{cc}
0&1\\
-1&0
\end{array}\right) \; .
\end{equation}
Inequality (\ref{bonfide}) is a useful way to 
express the Heisenberg uncertainty principle. \par 
In the following, we will make use of the Wigner 
quasi--probability representation
$W(x_{i},p_{i})$, defined as the Fourier transform of the 
symmetrically ordered characteristic function \cite{barnett}. 
In the Wigner phase space picture, the tensor product 
${\cal H}={\cal H}_{1}\otimes{\cal H}_{2}$ of the Hilbert 
spaces $H_{i}$'s of the two modes results in the direct sum 
$\Gamma=\Gamma_{1}\oplus\Gamma_{2}$ of the related phase spaces 
$\Gamma_{i}$'s. As a consequence of the Stone-Von Neumann theorem,
a symplectic transformation acting on the 
global phase space $\Gamma$ corresponds to a unitary operator acting on $\cal H$ 
\cite{simon87}. 
In what follows we will refer to a transformation $S_{l} = S_{1} \oplus S_{2}$, with each 
$S_{i} \in Sp_{(2,\mathbb R)}$ acting on 
$\Gamma_{i}$, as to a ``local symplectic operation''.
The corresponding unitary transformation is 
the ``local unitary transformation'' $U_{l}=
U_{1}\otimes U_{2}$, with each $U_{i}$ acting on ${\cal H}_{i}$.
Inequality~(\ref{bonfide})  
is then a
constraint on the $Sp_{(2,{\mathbb R})} \oplus Sp_{(2,{\mathbb R})}$ 
invariants \cite{simon}
\[
{\rm Det}\,\boldsymbol{\alpha}+\,{\rm Det} \,
\boldsymbol{\beta}+2
\,{\rm Det}\,\boldsymbol{\gamma}\le\frac{1}{4}
+ 4\,{\rm Det}\,\boldsymbol{\sigma}
\, .
\]\par
The Wigner function of a Gaussian state, 
written in terms of the phase space 
quadrature variables, reads
\begin{equation}
W(X)=\frac{\,{\rm e}^{-\frac{1}{2}X\boldsymbol{\sigma}^{-1}X^{T}}}{\pi
\sqrt{{\rm Det}\,\boldsymbol{\sigma}}}{\:,}
\label{wigner}
\end{equation}
where $X$ stands for the vector 
$(x_{1},p_{1},x_{2},p_{2})\in\Gamma$.
In general, the Wigner function transforms as a scalar 
under symplectic operations, 
while the covariance matrix $\boldsymbol{\sigma}$ transforms 
according to
$$
\boldsymbol{\sigma}\rightarrow 
S^{T}\boldsymbol{\sigma}S \; , \quad S \in Sp_{(4,\mathbb R)} \, .
$$ 
As it is well known, for any covariance 
matrix $\boldsymbol{\sigma}$ there exists a local 
canonical operation $S_{l}=S_{1}\oplus S_{2}$ which 
brings $\boldsymbol{\sigma}$ in the ``standard form'' 
$\boldsymbol{\sigma}_{sf}$ \cite{duan}
\begin{equation}
S_{l}^{T}\boldsymbol{\sigma}S_{l}=\boldsymbol{\sigma}_{sf}
\equiv \left(\begin{array}{cccc}
a&0&c_{1}&0\\
0&a&0&c_{2}\\
c_{1}&0&b&0\\
0&c_{2}&0&b
\end{array}\right) \; , \label{stform}
\end{equation}
where $a$, $b$, $c_{1}$, $c_{2}$ are determined 
by the four local symplectic 
invariants ${\rm Det} \,
\boldsymbol{\sigma}=(ab-c_{1}^2)(ab-c_{2}^2)$, 
${\rm Det}\,\boldsymbol{\alpha}=a^2$, ${\rm Det}
\,\boldsymbol{\beta}=b^2$, and
${\rm Det}\,\boldsymbol{\gamma}=c_{1}c_{2}$. 
Therefore, the coefficients of the standard form 
corresponding to any covariance matrix are unique 
(up to a common sign flip of the $c_{i}$'s).\par
\sirsection{Determination of the Von Neumann entropy}
To proceed, let us first note that
the purity $\mu$ (and therefore
the linear entropy $S_{L}$) of a Gaussian state
can be easily computed. In fact,
the trace of a product of operators corresponds to the 
integral of the product of their Wigner representations 
(when existing) over the whole phase space. Using the Wigner 
representation $W$ of ${\boldsymbol{\varrho}}$ and
taking into account the proper normalization factors,
for a $n$--mode Gaussian state we get
\begin{equation}
\mu= \frac{\pi}{2^{n}}
\int_{{\mathbb R}^{2n}}W^2\,{\rm d}^{n}x\,{\rm d}^{n}p =
\frac{1}{2^n \sqrt{\,{\rm Det}\,\boldsymbol{\sigma}}}\, .
\label{purezza}
\end{equation}
Eq.~(\ref{purezza}) implies that a Gaussian state $\boldsymbol{\sigma}$ 
is pure if and only if ${\rm Det}\,\boldsymbol{\sigma}=1/2^{2n}$.\par
For single--mode systems, the Von Neumann entropy 
can be easily computed as well.
Neglecting first moments,
any single--mode Gaussian state ${\boldsymbol{\varrho}}$ can in fact be 
written as  
\begin{equation}
{\boldsymbol{\varrho}} = 
S_{sm}(r,\varphi){\boldsymbol{\nu}}_{\bar{n}}S_{sm}^{\dag}(r,\varphi) 
\; ,
\label{genericonemode}
\end{equation}
where ${\boldsymbol{\nu}}_{\bar n}$ is a thermal state of 
mean photon number $\bar n$
\begin{equation}
{\boldsymbol{\nu}}_{\bar{n}}= \frac{1}{1+\bar{n}}\sum_{k=0}^{\infty}
\left(\frac{\bar{n}}{1+\bar{n}} \right)^k\: |k\rangle\langle k|
\: ,
\label{therma}
\end{equation}
and $S_{sm}(r,\varphi)=\exp(
\frac12 r \,{\rm e}^{-i2\varphi} a^{2} - 
\frac12 r \,{\rm e}^{i2\varphi} a^{\dag 2})$ 
is the single--mode squeezing operator. 
Being unitary, the latter does not affect the values
of the traces in Eqns.~(\ref{vneu})--(\ref{linea}), computed
on the diagonal density matrix ${\boldsymbol{\nu_{\bar{n}}}}$
given by Eq.~(\ref{therma}). One has then
\be
\mu({\boldsymbol{\varrho}}) = \frac{1}{2\bar n 
+1}=\frac{1}{2\sqrt{\,{\rm Det}\,
\boldsymbol{\sigma}}} \, , \label{pur1}
\ee
\bea
S_{V}({\boldsymbol{\varrho}}) &\hspace*{-.1cm}=&
\hspace*{-.1cm} \bar n \ln\left( 
\frac{\bar n +1}{\bar n}\right)+\ln(\bar n +1) \nonumber \\ 
&\hspace*{-.1cm}=&\hspace*{-.1cm} 
\frac{1-\mu}{2\mu}\ln\left(\frac{1+\mu}{1-\mu}\right)
- \ln\left(\frac{2\mu}{1+\mu}\right) \hspace*{-.1cm}. 
\label{vneu1}
\eea
Eq.~(\ref{vneu1}), first achieved in Ref.~\cite{agarwal71}, 
shows that for single--mode Gaussian states the
Von Neumann entropy is a monotonically increasing 
function of the linear entropy, so that 
$S_{V}$ and $S_{L}$ yield the same
characterization of mixedness. 
In fact, both $S_{V}$ and $S_{L}$ are fully 
determined by the same symplectic 
invariant $\,{\rm Det}\,\boldsymbol{\sigma}$.
As we will now see, this is no longer true 
for two--mode Gaussian states.\par
To find an expression for the Von Neumann entropy of a 
generic Gaussian state of a two--mode system, we must
find a general expression for the state analogous to 
that provided by Eq.~(\ref{genericonemode}) for a
single--mode system. Neglecting first moments, this
amounts to determine the most general  
parametrization of the covariance matrix, which is
provided by the following lemma.\par
\smallskip
\noindent{\bf Lemma 1}
{\it An arbitrary two--mode covariance matrix 
$\boldsymbol{\sigma}$ can be written as}
\begin{equation}
\boldsymbol{\sigma}=A^T 
\boldsymbol{\nu}_{n_{\mp}}A \; , 
\label{deco}
\end{equation}
where $\,\boldsymbol{\nu}_{n_{\mp}}=
\boldsymbol{\nu}_{n_{-}-1/2}
\oplus\boldsymbol{\nu}_{n_{+}-1/2}$ is the covariance
matrix of a tensor product of single--mode thermal states with average photon number 
$\bar n_{\mp}\equiv n_{\mp}-1/2$ in the two modes
\begin{equation}
\boldsymbol{\nu}_{n_{\mp}}
= \,{\rm diag}(n_{-},n_{-},
n_{+},n_{+})\; ,
\end{equation}
while
\begin{equation}
A = S_{loc}(r_{1},r_{2})R(\xi) S_{tm}(r)R(\eta) S_{l} 
\label{sytrasf}
\end{equation}
is a symplectic operation belonging to $Sp_{(4,\mathbb R)}$.
Transformation $A$ is made up by a local operation $S_{l}$, 
two rotations $R(\phi)$, with
\begin{equation}
R(\phi)=\left(\begin{array}{cccc}
\cos\phi&0&-\sin\phi&0\\
0&\cos\phi&0&-\sin\phi\\
\sin\phi&0&\cos\phi&0\\
0&\sin\phi&0&\cos\phi
\end{array} 
\right)\, , 
\label{rota}
\end{equation}
a two--mode squeezing $S_{tm}(r)=
\,{\rm diag}(\,{\rm e}^{r},\,{\rm e}^{-r},\,{\rm e}^{-r},\,{\rm e}^{r})$
and a local squeezing $S_{loc}(r_{1},r_{2})=
S_{sm}(r_{1},0)\oplus S_{sm}(r_{2},0)$, 
resulting from the direct product of
two single--mode squeezing operators 
with null phase. Note that $S_{tm}(r) = S_{loc}(r,-r)$,
so that the only global (nonlocal) operations in the 
decomposition of Eq.~(\ref{sytrasf}) are the two rotations.
We note that an equivalent decomposition has been recently
demonstrated for generic multimode pure Gaussian states \cite{botero}:
the authors have shown that for any decomposition of a multimode
pure Gaussian state with respect to a bipartite division of the
modes, the state can always be expressed as a product state involving
entangled two-mode squeezed states and single-mode local states
at each side.\par
\smallskip
\noindent{\em Proof.} 
In order to prove the statement expressed by Eq.~(\ref{deco}),
we consider the equivalent expression 
$A^{-1 T}\boldsymbol{\sigma}A^{-1}=
\boldsymbol{\nu}_{n_{\mp}}$ 
and show that it is realized by 
some $A$, $n_{-}$, $n_{+}$ 
for any given $\boldsymbol{\sigma}$.\\
First, we choose $S_{l}$ to 
bring $\boldsymbol{\sigma}$ to its standard form, given by Eq.~(\ref{stform}).
We then apply to $\boldsymbol{\sigma}_{sf}$
the rotated two--mode squeezing $R(\eta)^{-1}S_{tm}(r)^{-1}$, 
taking the covariance matrix to the form
$$
\left(\begin{array}{cccc}
s&0&0&0\\
0&m&0&c\\
0&0&s&0\\
0&c&0&n \end{array}
\right) \, , 
$$
which is convenient, due to the invariance of the 
submatrix $\boldsymbol{\delta}=\,{\rm diag}(s,s)$, 
see Eqns.~(\ref{minors}), under two--mode rotations
of the form Eq.~(\ref{rota}). 
The second rotation $R(\xi)^{-1}$ leaves $\boldsymbol{\delta}$ 
unchanged and can be chosen to make $c$ null, yielding a
state of the form
$$
\left(\begin{array}{cccc}
s&0&0&0\\
0&m'&0&0\\
0&0&s&0\\
0&0&0&n' \end{array}
\right) \, ,
$$
which can be finally put in the desired 
form $\boldsymbol{\nu}_{n_{-}-1/2}\oplus
\boldsymbol{\nu}_{n_{+}-1/2}$ by means of the 
local squeezing $S_{loc}(r_{1},r_{2})$. $\Box$\par
\smallskip
Lemma 1 introduces an equivalence relation on the set 
of Gaussian states, associating to any Gaussian state
$\boldsymbol{\sigma}$ a product of thermal states 
${\boldsymbol{\nu}}_{n_{\mp}}$, 
by means of the correspondence defined by Eq.~(\ref{deco}).
The quantities $n_{\mp}$ are known as 
the {\em symplectic eigenvalues} of 
$\boldsymbol{\sigma}$, while transformation $A$ 
performs a {\em symplectic diagonalization} \cite{holevo,werner,williamson}.
We note that the decomposition of the symplectic operation $A$ given by 
Eq.~(\ref{sytrasf}) is the particular two--mode case of the general 
decomposition of a symplectic operation \cite{agarwal}.\par 
Let us now focus on the quantity
\begin{equation}
\Delta ({\boldsymbol{\sigma}}) = \, {\rm Det}\,\boldsymbol{\alpha}
+ \, {\rm Det}\,
\boldsymbol{\beta}+ 2\,{\rm Det}\,\boldsymbol{\gamma} \, ,
\label{Delta}
\end{equation}
where $\boldsymbol{\alpha}$, $\boldsymbol{\beta}$, and 
$\boldsymbol{\gamma}$ are defined as in Eq.~(\ref{espre}).
We have:\par
\smallskip
\noindent{\bf Lemma 2.} 
{\em $\Delta ({\boldsymbol{\sigma}})$ is invariant under 
the action of the symplectic
transformation $A$ defined by} 
Eq.~(\ref{deco}).\par
\smallskip
\noindent{\em Proof.} 
$\Delta ({\boldsymbol{\sigma}})$ is clearly invariant 
under local operations, such as
$S_l$, $S_{loc}$ and $S_{tm}$. As 
for the non local rotations which enter in the 
definition of $A$, let us notice that they act 
on covariance matrices of the following form
$$
\tilde{\boldsymbol{\sigma}}=
\left(\begin{array}{cccc}
u&0&j&0\\
0&v&0&k\\
j&0&w&0\\
0&k&0&z \end{array}
\right) \, ,
$$
for which one has 
$$\Delta (\tilde{{\boldsymbol{\sigma}}})=\,{\rm Tr}\,
(\gr{\delta}\gr{\epsilon})\; .
$$
Such an expression is manifestly invariant under the action 
of identical rotations $R(\phi)$
on the submatrices $\boldsymbol{\delta}$ and 
$\boldsymbol{\gamma}$,
see Eqns.~(\ref{minors}) and (\ref{rota}). $\Box$\par
\smallskip
The quantity $\,{\rm Det}\,\boldsymbol{\sigma}$ is 
obviously invariant as well under the action 
of $A$ since, for any symplectic transformation $S$, 
one has $\,{\rm Det}\,S=1$.
Exploiting the invariance of $\,{\rm Det}\,\boldsymbol{\sigma}$ and 
$\Delta ({\boldsymbol{\sigma}})$ one can determine
the symplectic eigenvalues $n_{\mp}$ which characterize 
a generic Gaussian state $\boldsymbol{\sigma}$, according to 
Eq.~(\ref{deco})
\bea
{\rm Det}\,\boldsymbol{\sigma}
&=& \,{\rm Det}\,\boldsymbol{\nu}_{n_{\mp}}=
n_{-}^{2}n_{+}^{2} \, , \nonumber\\
&&\label{syst}\\
\Delta ({\boldsymbol{\sigma}}) &=&
\Delta ({\boldsymbol{\nu}}_{n_{\mp}})
=n_{-}^{2}+n_{+}^{2} \, . \nonumber
\eea
The solution of the system yields
\be
n_{\mp}(\boldsymbol{\sigma})  =  
\sqrt{\frac{\Delta ({\boldsymbol{\sigma}})\mp\sqrt{
\Delta ({\boldsymbol{\sigma}})^2 - 4 \,
{\rm Det}\,\boldsymbol{\sigma}}}{2}}\; .
\label{phther} 
\ee
Note that 
Ineq.~(\ref{bonfide})
is equivalent to 
\be
n_{\mp}\ge\frac12 \: , \label{sympheis}
\ee
whereas 
the necessary and sufficient criterion for a state to be pure reads 
$n_{-}=n_{+}=1/2$ (one can easily show that it
is equivalent to ${\rm Det}\,\boldsymbol{\sigma}=1/16$).\par
Knowledge of the symplectic eigenvalues and of the 
associated mean thermal photon numbers allows finally   
to determine the Von Neumann entropy $S_{V}(\boldsymbol{\sigma})$
of an arbitrary two--mode Gaussian state $\boldsymbol{\sigma}$. 
We have:\par
\smallskip
\noindent{\bf Proposition 1.} {\em The Von Neumann entropy 
$S_{V}(\boldsymbol{\sigma})$ of an arbitrary two--mode Gaussian state 
$\boldsymbol{\sigma}$ equals the one of the tensor product 
of thermal states $\boldsymbol{\nu}_{n_{\mp}}$, 
associated to $\boldsymbol{\sigma}$ 
via the correspondence established by} Eq.~(\ref{deco}),
{\em and its expression reads
\begin{equation}
S_{V}(\boldsymbol{\sigma})=f[n_{-}(\gr{\sigma})]+
f[n_{+}(\gr{\sigma})]
\; ,
\label{magicformula}
\end{equation} 
with $n_{\mp}$ given 
by} Eqns.~(\ref{phther}) {\em and}
\be
f(x)\equiv(x+\frac12)\ln(x +\frac12)
-(x-\frac12)\ln(x -\frac12) \, .
\label{magicfunc}
\ee
\smallskip
\noindent{\em Proof.} 
The symplectic operation $A$ described by Eq.~(\ref{sytrasf})
corresponds to a unitary transformation in the Hilbert space 
$\mathcal H$ which cannot affect the value of the trace 
appearing in the definition of $S_{V}$, according to Eq.~(\ref{vneu}). 
Therefore, exploiting Eq.~(\ref{vneu1}) and the 
additivity of the Von Neumann entropy for tensor product states, 
one obtains Eq.~(\ref{magicformula}). $\Box$\par
\smallskip
We have shown that the Von Neumann entropy 
of a two--mode Gaussian state $\boldsymbol{\sigma}$
depends on the two invariants 
$\Delta ({\boldsymbol{\sigma}})$ and 
${\rm Det}\,\boldsymbol{\sigma}$, 
whereas the purity of $\boldsymbol{\sigma}$ is completely determined by  
${\rm Det}\,\boldsymbol{\sigma}$ alone, just as in the
single--mode case. This implies that the hierarchy of mixedness 
established by the Von Neumann entropy on the set of 
Gaussian states differs,
in the two--mode case, from that induced by the 
linear entropy. States may exist with a 
given linear entropy, {\it i.e.}~with a given 
${\rm Det}\,\boldsymbol{\sigma}$, but with different 
Von Neumann entropies, {\it i.e.}~with different 
$\Delta ({\boldsymbol{\sigma}})$'s.
The Von Neumann entropy thus
provides a richer characterization of the 
state's lack of information.\par
\sirsection{Symplectic eigenvalues, mutual information and correlations}
The mutual information $I(\boldsymbol{\sigma})$ of 
a Gaussian state $\boldsymbol{\sigma}$ is defined as
\begin{equation}
I(\boldsymbol{\sigma}) = S_{V}(\boldsymbol{\sigma}_{1})+
S_{V}(\boldsymbol{\sigma}_{2}) - S_{V}(\boldsymbol{\sigma})\; ,
\label{defmi}
\end{equation}
where $\boldsymbol{\sigma}_{i}$ 
stands for the reduced single--mode state obtained 
by tracing over subsystem $j\neq i$.
Knowledge of  $S_{V}(\boldsymbol{\sigma})$ 
leads to the following:\par
\smallskip
\noindent {\bf Proposition 2.} {\em The mutual information 
$I(\boldsymbol{\sigma})$ of an arbitrary two--mode
Gaussian state is
\begin{equation}
I(\boldsymbol{\sigma})=f(a)+f(b)-f[n_{-}
(\boldsymbol{\sigma})]-f[n_{+}(\boldsymbol{\sigma})]\; ,
\label{misf}
\end{equation}
where $\,a=\sqrt{\,{\rm Det}\,\boldsymbol{\alpha}}\,$,
$\,b=\sqrt{\,{\rm Det}\,\boldsymbol{\beta}}\;$, and
f(x) is the same as in Eq.~(\ref{magicfunc})}.\par
\smallskip
\noindent{\em Proof.}  
Let us consider the reduction of $\boldsymbol{\sigma}$ to its 
standard form $\boldsymbol{\sigma}_{sf}$, defined by Eq.~(\ref{stform}).
The matrix elements $a$ and $b$ of $\boldsymbol{\sigma}_{sf}$ are 
easily recovered from a generic $\boldsymbol{\sigma}$, because 
${\rm Det}\,\boldsymbol{\alpha}=a^2$ and ${\rm Det}\,\boldsymbol{\beta}
=b^2$ are
$Sp_{(2,\mathbb R)}\oplus Sp_{(2,\mathbb R)}$ invariant.
Notice that, since either $S_{V}(\boldsymbol{\sigma})$ or the quantities 
$S_{V}(\boldsymbol{\sigma}_{i})$'s are invariant under 
local unitary operations, one has
$I(\boldsymbol{\sigma})=I(\boldsymbol{\sigma}_{sf})$.
Partial tracing of $\boldsymbol{\sigma}_{sf}$
over subsystem $i$ yields 
$\boldsymbol{\sigma}_{1}=\,{\rm diag}(a,a)$ and
$\boldsymbol{\sigma}_{2}=\,{\rm diag}(b,b)$, so that,
finally, Eq.~(\ref{vneu1}) and Proposition 1 lead 
to Eq.~(\ref{misf}). 
$\Box$\par
\smallskip
\noindent 
Notice that $a$ and $b$ constitute the symplectic spectra 
of, respectively, $\gr{\sigma}_{1}$ and $\gr{\sigma}_{2}$.
Eq.~(\ref{misf}) emphasizes the relevant role played by 
the symplectic eigenvalues 
$n_{\mp}(\boldsymbol{\sigma}_{sf})$ 
in determining the total amount of correlations contained
in a quantum state of a continuous variable system, in striking 
analogy to the role played by the symplectic eigenvalues of the 
partial transpose of $\boldsymbol{\sigma}_{sf}$ in characterizing 
the amount of quantum correlations \cite{giedke,wolf}. \par
To better clarify this point, 
let us consider a symmetric state $\boldsymbol{\sigma}_{sym}$, 
{\it i.e.}~a state whose standard form 
fulfills $a=b$, so that its mutual information
reads, according to Eq.~(\ref{misf})
\begin{equation}
{I(\boldsymbol{\sigma}_{sym})}
= 2f(a)-{f[n_{-}(\boldsymbol{\sigma}_{sym})]
-f[n_{+}(\boldsymbol{\sigma}_{sym})]}\; , 
\end{equation}
with symplectic eigenvalues $n_{\mp}=\sqrt{(a\mp c_{1})(a\mp c_{2})}$.
On the other hand, the symplectic eigenvalues of 
the partially transposed covariance matrix  
$\tilde{\boldsymbol{\sigma}}_{sym}$ (obtained from $\boldsymbol{\sigma}_{sym}$
by switching the sign of $c_{2}$, see \cite{simon}) 
are $\tilde{n}_{\mp}\equiv n_{\mp}(\tilde{\gr{\sigma}}_{sym})=
\sqrt{(a\mp c_{1})(a\pm c_{2})}$. 
In particular, for
an entangled state, the smallest eigenvalue 
is $\tilde{n}_{-}
=\sqrt{(a-|c_{1}|)(a-|c_{2}|)}$.
\footnote{If $\boldsymbol{\sigma}$ is entangled, then $\,{\rm Det}\,
\boldsymbol{\gamma}<0$, see Ref.~\cite{simon}.}\\
The symplectic eigenvalue 
$\tilde{n}_{-}$ 
encodes all
the information about the entanglement of the state, since 
the necessary and sufficient criterion for entanglement
reduces to $\tilde{n}_{-}<1/2$, 
while the entanglement of formation $E_{F}(\boldsymbol{\sigma}_{sym})$ 
reads 
\footnote{$E_{F}(\varrho)\equiv\min_{\{p_{i},|\psi_{i}\rangle\}}
\sum p_{i}E(|\psi_{i}\rangle\langle\psi_{i}|)$, 
where $E(|\psi\rangle\langle\psi|)$ is the entropy of entanglement 
of the pure state $|\psi\rangle$,
defined as the Von Neumann entropy of its reduced density matrix, 
and the $\min$ is taken over
all the pure states realization of 
$\varrho
= \sum p_{i}|\psi_{i}\rangle\langle\psi_{i}|$.}
\begin{equation}
E_{F}(\boldsymbol{\sigma}_{sym})
=\max\{0,g[\tilde{n}_{-}]\}\; , 
\end{equation}
with 
$
g(x) \equiv  \frac{(\frac12 + x)^{2}}{2x}
\ln \left(\frac{(\frac12 + x)^{2}}{2x}\right)
- \frac{(\frac12 - x)^{2}}{2x} 
\ln \left(\frac{(\frac12 - x)^{2}}{2x}\right)
$
(see \cite{giedke} for details); it
correctly reduces to 
$I(\boldsymbol{\sigma})/2$ for {\it pure} symmetric 
states. Even the quantification 
of quantum correlation provided by negativity \cite{werner},
which is computable also for non symmetric states, 
reduces for a two--mode Gaussian state $\gr{\sigma}$ to a 
simple function of $\tilde{n}_{-}({\gr{\sigma}})$.\par 
The dependence of entanglement on the eigenvalue $\tilde{n}_{-}({\gr{\sigma}})$ 
is due to the fact that the biggest eigenvalue of 
the partially transposed covariance matrix $\tilde{\gr{\sigma}}$ 
can be easily shown to fulfill Ineq.~(\ref{sympheis}). Thus
$\tilde{n}_{-}(\gr{\sigma})$ alone can be responsible
of the violation of the PPT (`positivity of the partial transpose')
criterion for separability \cite{simon}, which can be recast as
\be
\tilde{n}_{-}(\gr{\sigma}) \ge \frac12 \; .
\ee 
All the quantification of entanglement for two--mode Gaussian states 
available at present just quantify the violation of this inequality.
\par
We have extensively shown how both quantum and classical 
correlations of a Gaussian state 
are encoded in symplectic spectra of the global, 
reduced and partially transposed covariance matrices of the 
state.
\sirsection{Conclusions}
In conclusion, we have characterized 
mixedness and total correlations of two--mode Gaussian states 
by deriving their Von Neumann entropy
and mutual information. Comparing these quantities
with the entanglement of formation of symmetric states
shows that a crucial information about quantum  
and classical correlations lies in the 
symplectic eigenvalues of the covariance matrix and of its 
partial transpose. The problem is still left open of 
determining a fully satisfactory quantification 
of the {\it purely} quantum correlations in a 
general two--mode Gaussian state.\par
\ack
\noindent Financial support from INFM, INFN, and MIUR under
national project PRIN-COFIN 2002 is acknowledged.

\end{document}